\documentclass{PoS}

\title{Correlation of high energy neutrinos and gamma rays on the direction of Fermi Bubbles}

\ShortTitle{High energy neutrinos and gamma rays on the direction of Fermi Bubbles}

\author{\speaker{P. Alvarez-Hurtado} $^a$; N. Fraija$^a$, A. Galv\'{a}n$^a$ and A. Marinelli$^b$ \\$^a$ Instituto de Astronom\'{i}a, UNAM, A.P. 70-264, 04510 CDMX, M\'{e}xico\\$^b$ INFN, Pisa - Largo Bruno Pontecorvo 3, Pisa, Italy\\ E-mail: \email{palvarez@astro.unam.mx}, \email{nifraija@astro.unam.mx}, \email{agalvan@astro.unam.mx}, \email{antonio.marinelli@pi.infn.it}}

\abstract{We study the spatial correlation of astrophysical neutrinos detected by IceCube with the geometry of the two large globular structures located in the center of our Galaxy, known as Fermi Bubbles (FB). Using the Fermi-LAT data collected during 8 years and the upper limits derived by the High Altitude Water Cherenkov (HAWC) gamma-ray observatory, we use a hybrid (lepto-hadronic) model to investigate a possible correlation with the high-energy neutrinos in the direction of the Fermi Bubbles. We find that these events are possibly not associated with the Northern Bubble but do not dismiss a possible correlation with the Southern globular structure. We expect in the coming years to improve the gamma-ray observations through the Cerenkov Telescope Array (CTA) and the Southern Gamma-Ray Survey Observatory (SGSO) observatories to test a possible hadronic emission with the Southern  Bubble.}

\FullConference{36th International Cosmic Ray Conference -ICRC2019-\\July 24th - August 1st, 2019\\Madison, WI, U.S.A.}

\begin{document}

\section{Introduction}
Fermi bubbles, studied at high energies (0.1 to $\sim$300 \textrm{GeV}) through Fermi-LAT gamma-ray emission \cite{Dobler10, SuSlatyerFinkbeiner10}, are two giant structures located at 8-9 kpc.
They are symmetrically extended from the galactic plane with a characteristic non-thermal emission  globular shape from the center of our Galaxy. These Bubbles came up to be multi-messenger candidates with other multiples observations through microwaves \cite{Finkbeiner04}, X-rays \cite{Snowden97} and radio wavelengths \cite{Carretti13}.

Despite of the multiple observations that have been recorded, Fermi Bubbles emission remain a mystery. However, different scenarios have been proposed to explain the phenomenon responsible for the formation and the origin of the accelerated cosmic rays (CRs) involved in these.  According to magneto-hydrodynamic numerical simulations,  Mou et al. (2015) argued that the Fermi Bubble could have been formed  by powerful winds ($\sim 2\times10^{41}\,{\rm erg\,s^{-1}}$) launched to interstellar medium (ISM) and interacting with hot accretion flow in Sgr A*\cite{Mou15, CA11}. Lacki and Brian (2014) suggested that Fermi Bubbles could have been formed as a result of star formation  around the galactic center and molecular gas present in the CMZ by the interaction between outflow and ISM \cite{Lacki14}. Alternatively, another model, proposed by Cheng et al (2011), suggested that periodic capture processes of stars by the central super-massive black hole (SMBH) \cite{Cheng11, Cheng14} resulting in a tidal disruption event. Each one of the previous models can be cataloged by CR acceleration mechanism, that is, by hadronic, leptonic processes or by in situ acceleration.\\ Hadronic processes characterized by proton-proton (pp) interactions lead to the production of neutral pions through inelastic collisions between protons with a subsequent decays in two photons (e.g., see \cite{2012Fraija, 2014MNRAS}). In leptonic models, photons from cosmic microwave backgroud (CMB), ISFR and Infra-red (IR) are up-scattered through Inverse Compton (IC) effect by relativistic electrons producing gamma rays of high energies (e.g., see \cite{2017APh....89...14F, 2016ApJ...830...81F}). The in-situ acceleration scenario, it is proposed that CRs are accelerated by shocks or turbulence within the Bubbles, preferably near their edges\cite{Yang18}.\\ In 2013, the IceCube collaboration announced 5 events in the neighborhood of the Fermi Bubbles. Aartsen et al. \cite{Icecubecoll14} proposed that probable  connection between the gamma-ray spectrum and neutrinos detection. Since hadronic assumptions implies a neutrino counterpart, Fermi Bubbles were born as a possible extended neutrino sources.\\ Previous works \cite{LunardiniRazzaque12, Razzaque13} had made significant efforts searching for a correlation between the  Fermi-LAT data and the neutrino detection through hadronic models. In 2017, HAWC collaboration reported upper limits at very-high energies associated with the Northern Fermi Bubble.\\ In this work, we present a hybrid lepto-hadronic model in order to investigate a possible correlation among the Fermi-LAT emission, the VHE upper limits and  the high-energy neutrinos in the direction of the Fermi Bubbles. This work is organized as follows. In Section 2, we give the sample selection.  In section 3, we present the analysis method and discuss our results.  In section 4, we give a brief summary.
\section{Sample selection}
\begin{figure}[ht]
\centering
\subfigure[Neutrino high energy events]{\includegraphics[width=18pc]{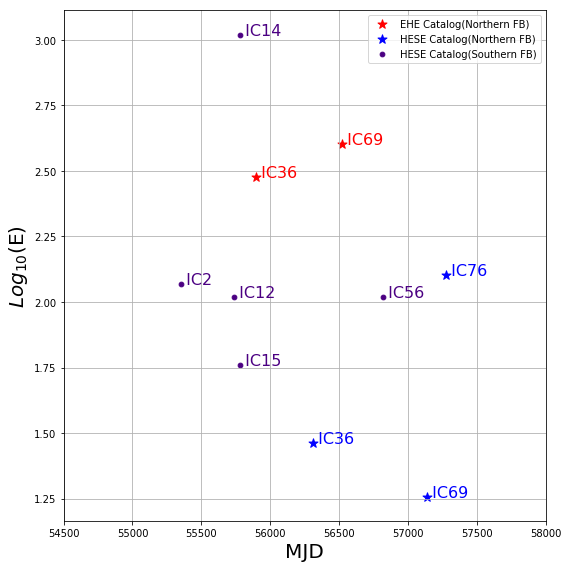}}
\subfigure[Neutrinos sample within FB]{\includegraphics[width=10pc]{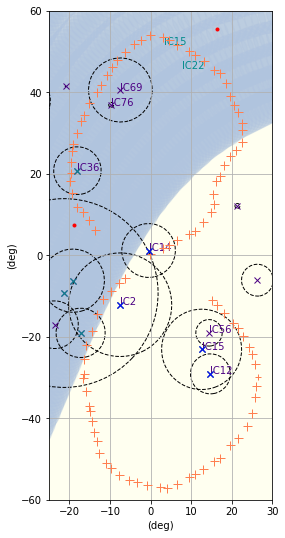}}
\caption{a) \label{mjd} Neutrinos events collected during 2101 days. b) \label{zoom} Location of 10 neutrino events spatially correlated with the Fermi Bubbles.}
\end{figure}

We use Fermi-LAT data collected during 8 years of observations. The data around bilobular structures is reprocessed using Pass 7 and Pass 8. Pass 7 dataset is used for both Bubbles while Pass 8 data is distinguished by each globular structure emission. We use the neutrino events reported  by IceCube telescope in the catalogues High-Energy Starting Events (HESE), the highest-energy Events (EHE) and those reported by the latest alerts before February 11th, 2019 (Figs.\ref{mjd},\ref{zoom}).\\ Due to poor angular resolution for the shower-like events, we only consider 10 neutrino events within Fermi Bubbles (2 neutrino tracks and 8 shower events with energies above 30 TeV, see Table \ref{tab1}). These neutrinos correspond to the events IC15, IC22, IC2, IC12, IC14, IC15, IC36, IC56, IC69 and IC76 (see Figure \ref{zoom}) respectively. These observations were performed during a lapse of 2101 days.\\ Within this sample the event IC14 deserve a special attention, with a deposited energy of $\sim1$ PeV at $|b|<10^\circ$ which is located near the galactic center with non-detection of low-energy neutrinos around this region.
%
\begin{center}
\begin{table}[ht]
\caption{Neutrino geometrically correlated with Fermi Bubbles.}
\label{tab1}
\resizebox{\textwidth}{!}{%
\begin{tabular}{@{}cccccccc@{}}
\toprule
Event ID & Energy (TeV) & $\delta E_{min}$ (TeV) & $\delta E_{max}$ (TeV) & Catalogue & Declination (deg) & Right Ascension (deg) & Topology \\ \midrule
IC2 & 117 & -14.60 & 15.4 & HESE & -28.00 & 282.6 & Shower \\
IC12 & 104.1 & -13.20 & 12.5 & HESE & -52.80 & 296.1 & Shower \\
IC14 & 1040.7 & -144.40 & 131.6 & HESE & -27.90 & 265.6 & Shower \\
IC15 & 57.5 & -7.80 & 8.3 & HESE & -49.70 & 287.3 & Shower \\
IC36 & 28.9 & -2.60 & 3 & HESE & -3.00 & 257.7 & Shower \\
IC56 & 104.2 & -10 & 9.70 & HESE & -50.10 & 280.5 & Shower \\
IC69 & 18 & -2 & 2.20 & HESE & 0.30 & 236.2 & Shower \\
IC76 & 126.3 & -12.7 & 12.00 & HESE & -0.40 & 240.2 & Shower \\
IC15 & 300 & -- & -- & EHE & 1.87 & 222.87 & Track \\
IC22 & 400 & -- & -- & EHE & -4.44 & 224.89 & Track \\ \bottomrule
\end{tabular}%
}
\end{table}
\end{center}


\section{Analysis Method: Searching for a possible correlation between gamma-rays and high-energy neutrinos}

We analyze first the viability of a hadronic pp-interaction model inferring the neutrino production within the timescale  $t_{acc}< t_{pp} \lesssim t_{esc}$ (e.g. see Ref. \cite{CA11, 2018F, 2018MNRASF, 2014ApJF, 2014MNRAS}. Since the timescales depend on the magnetic field, it is more viable the places where this field is $12\mu\,{\rm G}$ near to the Galactic Center region \cite{Razzaque13} and a few Gauss $\sim 2\mu\, {\rm G}$  at 8 kpc above Galaxy disk \cite{2019Shukurov}.
Taking into account the values reported, the timescales are $t_{acc}=0.28\,{\rm yr}$, $t_{pp}=3.52\times10^{9}\,{\rm yr}$ and $t_{esc}=3.92\times10^{9}\,{\rm yr}$ for a magnetic field of $12\mu $G in the GC and for $1\mu {\rm G}$,  $t_{acc}$=3.5 yr,  $t_{pp}=3.5\times10^{9}$ yr and $t_{esc}=3.2\times10^{9}$ yr.\\ Initially, we proposed a proton distribution model based on the analytically approach of Kelner et al (2006), which is governed by a power law with an exponential cutoff of the form: 
\begin{equation}
\label{ec1}
J_p(E_p)={A_o}_{\gamma} E_p^{-\alpha_{\gamma}}exp [- ( {E_p}/{E_o}_{\gamma} )^{\beta} ]\,. 
\end{equation}
Here, it is adjusted to the gamma-ray spectrum using the parameters ${A_o}_{\gamma}$, ${\alpha}_{\gamma}$ and ${E_o}_{\gamma}$ (Table \ref{tab2}).  For the Nothern Bubble, we use, in addition, the VHE upper limits reported by HAWC observatory. These are in the range of  $3\times10^{-7}\, \textrm{GeV} \textrm{cm}^{-2}\, \textrm{s}^{-1}\,\textrm{sr}^{-1}$ and $4\times10^{-8}\, \textrm{GeV}\, \textrm{cm}^{-2}\, \textrm{s}^{-1}\,\textrm{sr}^{-1}$ in the energy range from 1.2 at 123.7 TeV \cite{Abeysekara17}.

The maximum energy that protons can be accelerated is obtained through the timescales. Given the neutrinos events, the neutrino flux can be estimated as \cite{Razzaque13}:
\begin{equation}
 \phi_{\nu} E^{2} =\frac{ n\, E_{\nu}}{ 4\pi\, A\, t} .
 \label{fluxneutrino}
\end{equation}
with $\rm{n}$ is the number of events, $E_\nu$ is the events energy in GeV, $\rm{A}$ is the effective area  in $\rm{cm}^{2}$ and $\rm{t}$ corresponds to experiment livetime for the events collection in seconds.
\begin{table}
\centering
\caption{Parameters for the construction of gamma ray spectrum fits in the hadronic model.}
\label{tab2}
\resizebox{0.85\textwidth}{!}{
\begin{tabular}{@{}ccccccccc@{}}
\toprule
\multicolumn{4}{c}{\textbf{Northern}}            & \multicolumn{1}{l}{} & \multicolumn{4}{c}{\textbf{Southern} }             \\ \midrule
\textbf{Fit} & \textbf{${E_o}_{\gamma}$ (PeV)} & \textbf{ ${\alpha}_{\gamma}$} & \textbf{${A_o}_{\gamma}$ (GeV)} &                      & \textbf{Fit} & \textbf{${E_o}_{\gamma}$ (PeV)} &\textbf{ ${\alpha}_{\gamma}$}  & \textbf{${A_o}_{\gamma}$ (GeV)}  \\
N1  & $2$ & $2.15$ & $6.5\times 10^{55}$  &                      & S1  & $2.5$ & $2$ & $8.74\times 10^{56}$  \\
N2  & $1.7$ & $2$ & $3.87\times 10^{57}$  &                      & S2  & $3$ & $2.2$ & $2\times 10^{58}$ \\ \hline
\end{tabular}
}
\end{table}
Using eq. (\ref{fluxneutrino}) we calculate the flux associated of our  sample through a statistical distribution in a total range from 15 TeV to 4 PeV. The data uncertainties are associated with the energy of the neutrino events in each Bubble solid angle. The corresponding area was determined by the response function of the IceCube for an observational time of 2101 days. 

Due to poor evidence of an association between these 10 neutrinos and gamma-rays generated mainly by a population of protons in both Bubbles, we propose a hybrid model consisting of a leptonic and hadronic scenario to explain the origin of the emission to VHE. In the leptonic scenario, electrons are accelerated with protons and are described by a simple power-law $dN_e/dE \sim E_e^{-2.2}$. We assume that the high-energy photons are generated by the IC scatterings from CMB, IR and SL photons.

The dominant supply of low-energy photons is transferred through CMB photons with an energy density of $4\times10^{-13}$ erg cm$^{-3}$. Other seed photon field also contributes to IC scatterings although less significantly. In this case, the energy densities for the stellar radiation and infrared photons are $2.3\times10^{-13} $erg cm$^{-3}$ and $0.24\times10^{-14}$ erg cm$^{-3}$, respectively \cite{2011Guo}.\\ In the hadronic scenario, we use the power-law with exponential cutoff give in eq.(\ref{ec1}). The best-fit values are reported in Table 2. In this table can be seen that the spectral index is close to $\sim 2$ for both Bubbles, and  the cut-off energy varies from a minimum of 1.7 TeV to 3 PeV.

\begin{figure}[t]
\centering
\includegraphics[width=27pc]{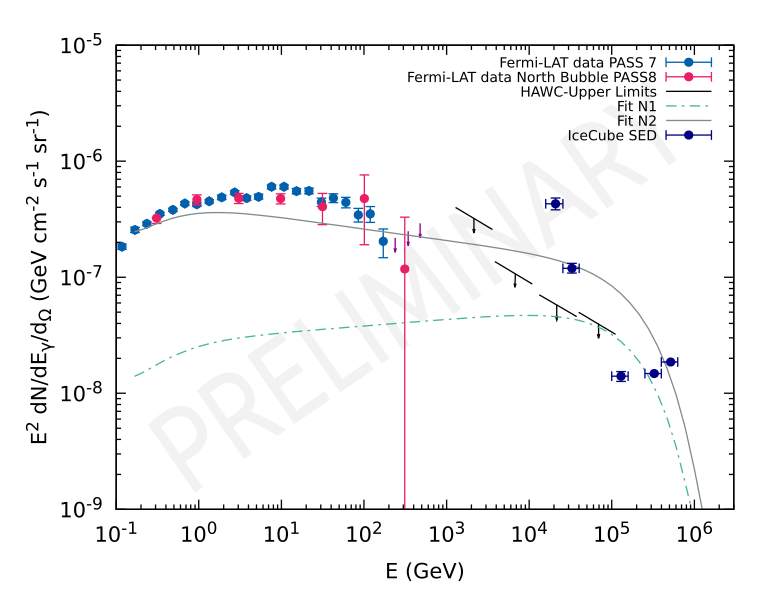}
\caption{\label{Northern}Lepto-hadronic spectrum for the Northern bubble.}
\end{figure}

\begin{figure}[th]
\centering
\includegraphics[width=27pc]{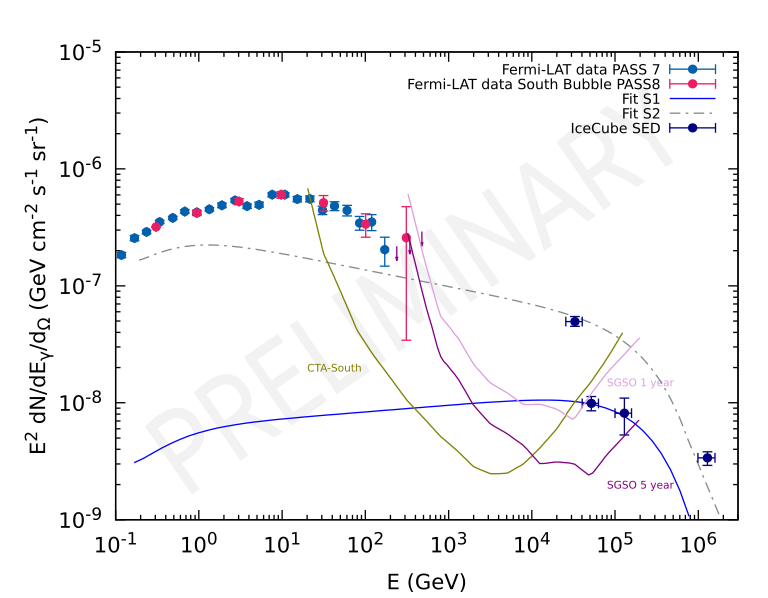}
\caption{\label{Southern}Lepto-hadronic spectrum for the Southern bubble.}
\end{figure}

Figures \ref{Northern} and \ref{Southern} show the Fermi-LAT data, the IceCube fluxes and the lepto-hadronic model for the Northern and Southern Bubbles, respectively. In addition, the VHE upper limits of the HAWC observatory were placed for the Northern Bubble, and the sensitivities of future experiments such as  \textit{Cherenkov Telescope Array} (CTA) and \textit{Southern Gamma-ray Survey Observatory} (SGSO) were added for the Southern Bubble.\\ Figure \ref{Northern} shows two different fits for the Northern Bubble, one can describe the Fermi-LAT data except above the VHE upper limits, the other that cannot describe the Fermi-LAT data but can be used to describe the neutrino flux. In both cases, the gamma-ray and neutrino fluxes are hardly correlated. Therefore we conclude that the high-energy neutrinos are not associated with the Fermi Northern Bubble through a single model. 
The production of HESE events reconstructed inside the Northern Fermi Bubbles solid angle cannot be simply connected to the observed gamma-ray emission through a single model. Therefore suggesting a different origin for the sample of the HESE events selected in this region.\\ Figure \ref{Southern} shows again two different fits. The fit S1 could interpret both the gamma-rays and neutrinos and the fit S2, only the neutrinos for the Southern Bubble. For the Southern Bubble case, despite the absence of VHE upper limits, the fits may poorly explain the emission in the region of energies above $\gtrsim 10$ TeV we considered the expected sensitivities of CTA and SGSO telescopes.

\section{Conclusions and Outlook}

A hybrid model was proposed in this work to correlate the high-energy photons detected by Fermi-LAT and high-energy neutrino events detected by IceCube in the Northern and Southern Fermi Bubble's direction. We have considered the neutrinos reported by IceCube until February 11th, 2019 and the VHE upper limits derived by HAWC observatory.\\ The contribution of the models is presented in two parts, the first part is obtained assuming a leptonic mechanism and the second is associated with a subdominant hadronic component.\\ According to our estimates, the number of events expected during 2101 days indicates that the probability to associate the neutrinos with the Fermi Bubbles is low, although more studies are still required, with new observations. In particular, a more detailed study of the event IC14 detected in the vicinity of the SMBH Sgr A* is required (1 PeV).\\ From the analysis carried out, it is suggested that the events recorded by IceCube in the Northern bubble are cannot be easily correlated to the accelerated cosmic rays that can produce the gamma rays observed by Fermi-LAT. However, we can not give a similar conclusion for the Southern Bubble due to the limited field of view of the current Cherenkov detectors.\\ We expect in the coming years, the beginning of observations of gamma rays up to $\sim$300 GeV in observatories such as CTA and SGSO able to improve the statistics of gamma rays observed from this region of our Galaxy.

\section{Acknowledgement}
We acknowledge the financial support from UNAM-DGAPA-PAPIIT through Grant IA 102019. \\ 

\end{document}